# Properties of Pt Schottky Type Contacts On High-Resistivity CdZnTe Detectors

Aleksey E. Bolotnikov[*], Steven E. Boggs, C. M. Hubert Chen, Walter R. Cook, Fiona A. Harrison, and Stephen M. Schindler

California Institute of Technology, Pasadena, CA 91125

**Abstract**

In this paper we present studies of the *I-V* characteristics of CdZnTe detectors with Pt contacts fabricated from high-resistivity single crystals grown by the high-pressure Brigman process. We have analyzed the experimental *I-V* curves using a model that approximates the CZT detector as a system consisting of a reversed Schottky contact in series with the bulk resistance. Least square fits to the experimental data yield 0.78-0.79 eV for the Pt-CZT Schottky barrier height, and <20 V for the voltage required to deplete a 2 mm thick CZT detector. We demonstrate that at high bias the thermionic current over the Schottky barrier, the height of which is reduced due to an interfacial layer between the contact and CZT material, controls the leakage current of the detectors. In many cases the dark current is not determined by the resistivity of the bulk material, but rather the properties of the contacts; namely by the interfacial layer between the contact and CZT material.

**Keywords**: X-ray astrophysics, CdZnTe pixel detectors, *I-V* curve measurements

## 1. Introduction

The dark current is a critical parameter that for many configurations can be the primary factor limiting the energy resolution of CdZnTe (CZT) detectors. In the course of developing a focal plane detector for the balloon-borne High-Energy Focusing Telescope (HEFT) [1], we carried out routine measurements of the dark current characteristics for a large number of CZT pixel detectors of a specific pixel contact design. Our detector anode pattern includes very thin strips (a grid) between the pixel contacts, held at a small negative potential. The real purpose of this biased grid is to enhance the charge collection near the surface between pixel contacts. However, for the dark current measurements we can ground the grid, so that it serves as a guard ring to eliminate surface leakage currents, allowing accurate measurement of both surface and bulk leakage.

We tested a large number of CZT detectors, measuring the surface and bulk *I-V* curves over a wide voltage range. We found large variations in the shapes and nominal surface dark currents for different detectors, as well as for different pixels of the same detector. This is the case even for detectors where the specific bulk resistivity, as defined by approximating *I-V* curves to Ohm's law at very low bias, <0.5 V, varies only by 20-40%. In some detectors, the measured *I-V* characteristics also resemble a simple Ohm's law at higher bias. The specific resistivity, evaluated by fitting the data from a high voltage region, significantly exceeds the upper limit established for the CZT material used in these measurements, $\sim 5 \times 10^{10}$ Ohm-cm at 26 C [2].

To understand these experimental leakage current measurements, we modeled the CZT detector as a metal-semiconductor-metal (MSM) system with two back-to-back Schottky barriers. Two simplified treatments have been previously applied to such a system: Sze et al. [3] used the thermionic-limited approximation of the Schottky barrier, and Cisneros et al. [4] treated the barrier in the diffusion-limited approximation. Neither of these approaches could explain our dark current measurements. In our previous work [5] we briefly pointed out that the experimentally measured currents were considerably smaller than the saturation thermionic current expected for the Pt-CZT, and the measured *I-V* curves differed in shape from the diffusion-limited current expected for two back-to-back Schottky barriers.

Although the models described above are over-simplified, we also cannot explain our *I-V* curve measurements over the full voltage range even with a more general treatment of an MSM system. Crowell and Sze [6] demonstrated that the thermionic- and diffusion-limited models are not independent, but are in fact limiting cases of a more general thermionic-diffusion theory. Using this theory we can reproduce the measured *I-V* curves at a low voltages (in some cases up to 100 V), but at high voltages the measured current increases much faster than predicted by the theory. One might expect that the discrepancy could be explained by tunneling across the interface (normally the dominant current in highly doped semiconductors at low temperatures). For this to be the case, our measurements show that tunneling would have to start to contribute at ~50 V (for a 2 mm thick detector). At this low voltage, the total current across the CZT is much less than the expected saturation thermionic current (see Eq. (21) and discussion below). The tunneling component should, however, become important at much higher biases, where the thermionic emission component is close to its saturation limit (~500 V).

---

[*] Correspondence: Email: bolotnik@srl.caltech.edu; Telephone: 626-395-4488

We find that to explain the shape of our *I-V* curves, we must assume the existence of a very thin (10-100 nm) insulating layer (residual oxide layer) between the contact and the semiconductor material, which could be formed before or after metal contacts are deposited [7-12]. To include the effects of an interfacial layer in the Schottky barrier model, Wu [14] developed a combined interfacial layer-thermionic-diffusion (ITD) model. We show that adopting this ITD model allows us to accurately fit the experimental data without considering any other possible current components (such as tunneling, or generation recombination currents). We demonstrate that by taking into account the interfacial layer we can explain the full variety of measured *I-V* curves, and by fitting the data we can obtain for each detector a consistent set of parameters that characterize the Schottky barrier and CZT material.

## 2. Theoretical background and fitting algorithm

This section briefly describes the theoretical model of the Schottky barrier with a thin interfacial layer, as applied to the MSM system, which we employ in our analysis. For details we refer to the original work by Sze et al. [3], Sze [15], Cisneros et al. [4], Wu [14], and Cohen et al. [16]. From the mathematical point of view a Pt-CZT-Pt MSM system is rather complicated. Fortunately, because of the high bulk resistivity of semi-insulating material such as CZT, we can make some simplifications. The series resistance of the undepleted bulk material is much higher than the resistance of the forward-biased Schottky barrier at the anode, and the width of its depleted layer is much smaller than the total thickness of the CZT crystal. We can therefore neglect the effect of the anode contact. This simplification allows us to treat a CZT detector as a metal-semiconductor system consisting of a reversed-biased Schottky barrier at the cathode coupled to the series resistance of the bulk. The band diagram of this system is shown in Fig. 1.

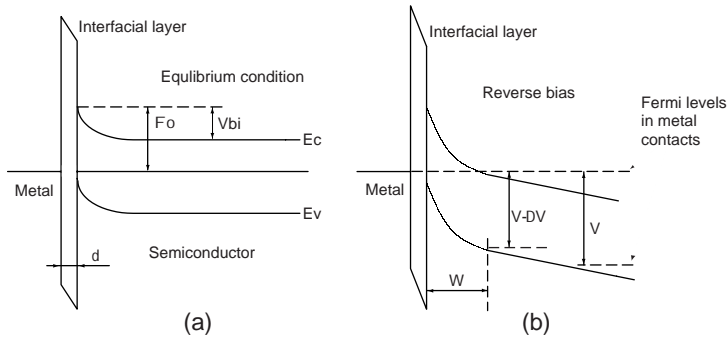

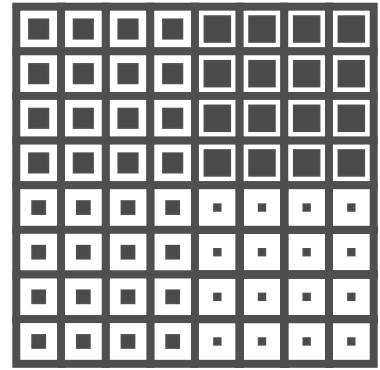

Figure 1. Schottky contact with interfacial layer (a) unbiased and (b) reverse biased.

Figure 2. Contact pattern with a focusing grid.

The detectors we have studied have rectangular pixel contacts surrounded by a grid on the anode side (see Fig. 2.) and a monolithic contact on the cathode side. We treat this as a one-dimensional system, where the electric field is uniform in both *X* and *Y* directions. In the Schottky-depleted-layer approximation, if a small negative voltage, -*V* (*V*>0) is applied to the cathode, the electric field distribution, *U(z),* inside both the depleted and undepleted regions of the detector can be written as:

$$U(z)=(eN_D/2\varepsilon)(z-W)^2 - E_A(z-W) + \Delta V, \qquad 0<z<W \qquad (1)$$

and

$$U(z)=E_A(W-z)+\Delta V, \qquad W<z<L. \qquad (2)$$

In the above equations, $W$ is the width of the depleted layer, $L$ is the thickness of the CZT crystal, $E_A$ is the electric field strength inside the undepleted bulk (same as at the anode), $\varepsilon$ is the permittivity of CZT, $e$ is the electron charge, $N_D$ is the concentration of the ionized donor centers, and $-\Delta V$ ($\Delta V>0$) is the potential at the edge of the depleted layer ($\Delta V=(L-W)E_A$). Using the boundary condition at the cathode and at the edge of the depleted layer, one can find the width of the depleted layer from:

$$V+V_{bi}=(eN_D/2\varepsilon)W^2+\Delta V, \tag{3}$$

where $V_{bi}$ is the built-in voltage or diffusion potential at the cathode (see Fig. 1). From this equation, $W$ can be calculated if $E_A$ or $\Delta V$ is known. If $W/L<<1$ and $\{(eN_D/2\varepsilon)W^2-V_{bi}\}/V<<1$ then $E_A \equiv \Delta V/(L-W)=V/L$, and $I\sim V$, i.e. at small applied biases the current follows Ohm's law. The voltage $V_{RT}$ required to deplete the whole volume of the crystal, defined as the *reach-through* voltage, is given by:

$$V_{RT}=(eN_D/2\varepsilon)L^2+E_0L-V_{bi} \tag{4}$$

where $E_0$ is the electric field strength at the anode when the cathode is at $V_{RT}$, i.e. $E_0=E_A(V_{RT})$. Notice, that when the bulk resistance is neglected, $E_0=0$, and Eq. (4) becomes the standard expression for the *flat-band* voltage–a parameter usually defined to characterize the back-to-back barrier system [3,4]. For applied voltages higher than $V_{RT}$:

$$U(x)=V_{RT}(z-L)^2/2L+(z-L)V/L. \tag{5}$$

Correspondingly, the electric-field strength at the cathode $E_C$–the parameter which we will need for further calculations–is given by:

$$E_C(V)=(eN_D/\varepsilon)W+E_A, \qquad V<V_{RT}, \tag{6}$$

and

$$E_C(V)=(V_{RT}+V)/L-E_0, \qquad V>V_{RT} \tag{7}$$

In the combined ITD model, the reverse current, $I_R$ (A/cm$^2$), over the barrier at the cathode is expressed as [13]:

$$I_R=\{\vartheta_n A^* T^2/(1+\vartheta_n V_R/V_D)\}exp(-\Phi_R/V_{TH})(1-exp\{-(V-R_S I_R)/V_{TH}\}), \tag{8}$$

where $A^*$ is the effective Richardson constant, $T$ is the temperature, $V_R$ is the thermal velocity, $\vartheta_n$ is the transmission coefficient through the oxide layer, $R_S$ is the series resistance of the bulk, and $V_{TH}=kT/e$. $V_D$ is an effective diffusion velocity [11,14] that can be calculated analytically if Eq. (1) is used to approximate the field distribution in the depleted layer. In this case, $V_D$ is simply the electron drift velocity at the cathode, namely:

$$V_D=\mu E_C, \tag{9}$$

where $\mu$ is the electron mobility ($\mu=1000$ cm$^2$/Vs). The effective Richardson constant is related to the thermal velocity $V_R$ by:

$$A^* T^2 \equiv V_R N_C, \tag{10}$$

where $N_C$ is the effective density of the states in the conduction band given by:

$$N_C=2(2\pi m_0 kT/h^2)^{3/2} \tag{11}$$

The Schottky barrier height, $\Phi_R$, is a function of the applied voltage and reflects the barrier lowering due to the voltage drop across the oxide layer. Again, following Wu [14], we assume that $\Phi_R$ depends linearly on the applied voltage (the barrier lowering due to image-force is negligible in our case) given by:

$$\Phi_R=\Phi_0-(1-1/n_0)V, \tag{12}$$

where $\Phi_0$ is the barrier height under thermal equilibrium conditions, with

$$1/n_0=\varepsilon_i/(\varepsilon_i+e2\delta D_S). \tag{13}$$

Here $\varepsilon_I$ and $\delta$ are the permitivity and thickness of the interfacial layer, and $D_S$ is the density of surface states per unit energy and area.

The series resistance of the undepleted layer can be expressed as:

$$R_S=(L-W)/eN\mu, \tag{14}$$

where $N$ is the free electron concentration (we assume that CZT is an n-type semiconductor). Substituting Eq. (14) into Eq. (8) and using Eq. (5) for $W$ and Eqs. (6,7) for $E_C$ we can numerically calculate the I-V dependence for the current across the whole system. The above equations contain too many free parameters, and the information contained in a single I-V curve is obviously insufficient to obtain the parameters from a fitting procedure. Our primary goal, however, is not to evaluate all these parameters explicitly, but to demonstrate that by assuming reasonable values for these parameters, the measured I-V curves can be explained with the ITD model.

The effective Richardson constant can be calculated as $A^*=120(m^*/m_0)$ (in A-cm$^{-2}$K$^{-2}$), where $m^*$ and $m_0$ are the effective and free electron masses. Since the ratio $m^*/m_0$ for ZnTe and CdTe are 0.11 [17] and 0.09 [10], respectively, we assume for CZT a similar ratio of 0.1. Thus, $A^*=12$ A-cm$^{-2}$K$^{-2}$. $N$ can be evaluated from Eq. (14) after fitting the I-V curve at low voltages where the dependence follows Ohm's law ($W<<L$ and $\Delta V=V$). For the typical intrinsic bulk resistivity of $3\times10^{10}$ Ohm–cm, $N=2.5\times10^5$ cm$^{-3}$. The limits for the potential barrier height $\Phi_0$ can be found from results obtained for Pt-CdTe and Au-CdTe systems (see e.g. Refs. [10,18]) where $0.7<\Phi_0<0.9$. As for $V_{RT}$, $\vartheta_n$ and $n_0$, these parameters depend on the contact fabrication process, and have to be found by fitting the I-V curves.

In the high voltage region, where the crystal is fully depleted ($R_S=0$) Eq. (8) can be simplified:

$$I_R=\{C_0/(1+C_1/(V_{RT}+V-E_0L))\}exp(C_2V). \tag{15}$$

Here

$$C_0=\vartheta_n A^*T^2 exp(-\Phi_0/V_{TH}), \tag{16}$$

$$C_1=\vartheta_n LV_R/\mu, \tag{17}$$

and

$$C_2=1-1/n_0. \tag{18}$$

If the effect of interfacial layer is negligible, $C_2=0$ and $\vartheta_n=1$. From Eq. (10) one can find the following expression for the ratio $C_0/C_1$:

$$C_0/C_1=(N_C\mu/L)exp(-\Phi_0/V_{TH}), \tag{19}$$

which allows us to estimate the potential barrier $\Phi_0$.

To fit the experimental data, we first assume that the parameters $V_{RT}$ and $C_2$ are known, and apply Eq. (15) to fit the I-V curve for the voltages above $V_{RT}$ (high enough that $E_0L/V_{RT}<<1$). We then evaluate the parameters $C_0$ and $C_1$ and use these to calculate the potential barrier height, $\Phi_0$, from Eq. (19), and the $\vartheta_n V_R$ product from Eq. (17). $\vartheta_n$ and $V_R$ cannot be evaluated separately, however, since we assumed that $A^*$ is known and equal to 12 A-cm$^{-2}$K$^{-2}$, then from Eq. (10): $V_R=8.5\times10^6$ cm/s. We then find $E_0$ by solving Eqs. (7) and (8) with $V=V_{RT}$, and $R_S=0$. This allowed us to calculate $N_D$ from Eq. (4). Finally, we minimized the $\chi^2(V_{RT},C_2)$ function, given by:

$$\chi^2(V_{RT},C_2)=\Sigma\{(I_{CAL}-I_{MEAS})/\sigma\}^2, \tag{20}$$

to obtain estimates for $V_{RT}$ and $C_2$. Note that for $V<V_{RT}$, we solve Eq. (3) and Eq. (8) numerically to calculate $W$, $I$ and $E_A$ for each applied voltage $V$.

## 3. Experimental setup

We measured *I-V* dependencies using a probe stage with a GPIB-controlled HP 3458A multimeter and a EDC 521 DC calibrator. All measurements were taken at a steady state current condition. Because of the large number of deep traps in the CZT material, it can take several minutes or even hours to reach equilibrium between free and trapped charge. These measurements are therefore very time consuming, and we use a computer-controlled setup.

To reduce the waiting time before equilibrium is reached, we varied the bias on the cathode in small steps. After each step, we paused for several minutes before taking 10-20 sequential measurements of the current, separated in time by 1-min intervals. This sequence of data points allows us to verify that equilibrium has been actually achieved, and also to improve the accuracy of the measurements. We took the majority of measurements at room temperature, (26 +/-1 C). For one detector, we varied the temperature from 17 to 70 C. We place the detector on a hot-plate, covered by a super-insulating screen. During the measurement the temperature stability was +/-0.5 C, monitored with a thermocouple (accuracy +/-0.1 C) attached to the hot-plate in close proximity to the detector.

We used four groups of CZT pixel detectors, fabricated by eV-Products over a two-year period. The first two groups, labeled D1 and D2, were fabricated (to the best of our knowledge) from different slices of the same ingot, and we therefore expect them to have similar performance. Detectors from groups D1 and D4 are 12x12x2 mm CZT crystals, each with a single 8x8 mm contact enclosed inside a guard ring on one side, and a monolithic contact on the opposite side. The gap between the contact and the guard ring is 0.2 mm. The detectors from group D2 are 8x8x2 mm single crystals, with four patterns of 8x8 pixel arrays (see Fig. 2). The physical size of a pixel is 650 by 680 µm. 50 µm wide orthogonal strips are placed between the pixel contacts. Each pixel from a pattern has the same gap between the contact and the grid, which varied between 100 and 250 µm from pattern to pattern. Finally, the detectors from the third group, D3, are 7.1x7.1x1.7 mm CZT crystals fabricated from a different ingot. The D3 detectors have pixel patterns similar to D2, except the pixel size is 400 by 400 µm, and the gaps between contacts and grid are 50 and 75 µm, with a 25 µm grid width.

We typically took the measurements from -100 V to +100 V between contacts and cathode, but for some detectors we increased the maximum applied voltage up to 1 kV. We eliminated the leakage current flowing over the side surfaces of the detector by using a guard ring.

## 4. Results and discussion

Figs. 3-5 show the *I-V* characteristics for the three groups of detectors, measured for bias voltage of -100 to +100 V. In these plots the currents are normalized to the *effective* area of the pixel contact (i.e. to the geometrical area with boundaries

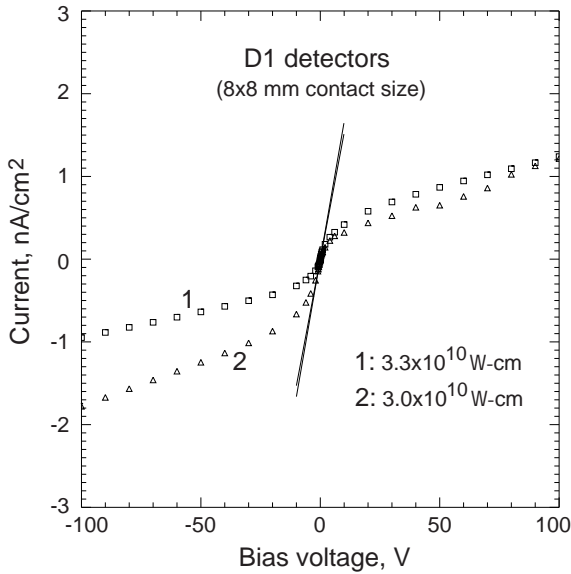

Figure 3. *I-V* characteristic measured for two D1 detectors. The contact size is 8x8 mm; the gap between the contact and the guard ring is 200 µm; the effective contact area used to normalize the current is 0.672 cm$^2$.

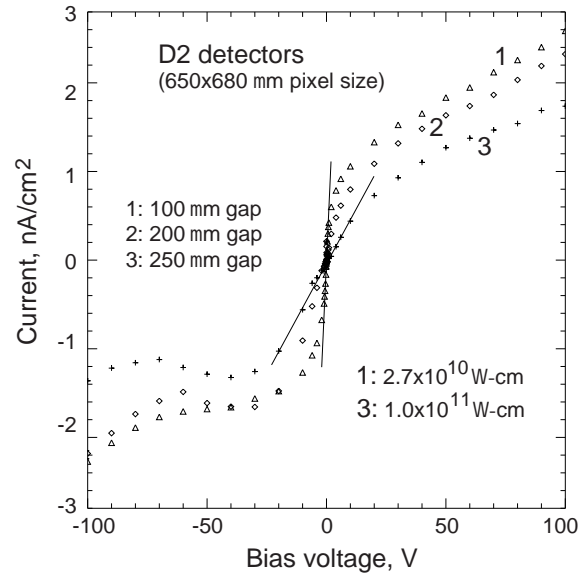

Figure 4. *I-V* characteristic measured for three D2 detectors; the pixel size is 650x680 µm; the gaps between the contact and the guard ring (grid) in µm are: (1) 100, (2) 200, (3) 250. The effective contact areas used to normalize the current in cm$^2$ are (1) 0.00264, (2) 0.00171, and (3) 0.00132.

in the middle of the gap between contact and grid). This approximation works only for small gaps. Fig. 3 shows the curves measured for the two pixels of one of the D1 (large contact) detectors. The shape of the curves clearly indicates the existence of Schottky barriers on the anode and cathode sides of the detectors. At low applied biases (<1 V) the *I-V* curves follow Ohm's law, with the slopes corresponding to a specific resistivity of $2.9 \times 10^{10}$ and $2.2 \times 10^{10}$ Ohm-cm for detectors D1 and D2 respectively. These are typical values for high-resistivity CZT material grown by eV-Products. As the voltage increases, the linear slope starts to change. When the absolute voltage is between 1 and 50 V, the I-V relations again becomes close to a linear law, but with a slope several times smaller.

We observed similar behavior for the D2 detectors (small pixel contacts). Fig. 4 shows a set of curves measured for several different size pixels. Only the positive branches of the *I-V* curves (cathode is positive biased) exhibit the described behavior. The negative branches seem to be affected by the surface conductance in the gap between the guard ring and the contact, and show a slightly different behavior. Here the current reaches a local maximum at around -25 V and then decreases and starts rising again (negative dynamic resistance). This asymmetry of the positive and negative branches indicates that the CZT crystal is a n-type. Indeed, when a *positive* bias is applied to the cathode, a depleted layer starts to expand from a pixel contact (for an n-type CZT) toward the cathode and along the surface into the gap between the contact and the guard ring (a fringe effect). Effectively, this increases the area of the contact until the whole area along the surface becomes depleted. This happens at relatively low biases, for which the measured current is still bulk resistance limited. At positive biases on the cathode, the fringe effect does not show up in the *I-V* curves. However, when a *negative* bias is applied to the cathode, the depleted layer starts to grow from the cathode, reaching the anode side (pixel contacts first) when the bulk resistance becomes negligible. At high absolute bias (>100 V), the negative and positive branches of *I-V* curves behave similarly. Because of the surface effects, we cannot estimate the specific resistivity of the CZT for the pixels with large gaps between contacts and grid. For example, the bulk resistivity evaluated for a 250 µm gap pixel was greater that $10^{11}$ Ohm-cm (curve 3 in Fig. 4) which is obviously an unrealistic value. For several *I-V* curves, measured for pixels of both the D1 and D2 detectors, we extended the maximum applied bias up to +/-400 V. These measurements revealed that above 100-150 V, the linear portion of the *I-V* curves is followed by an exponential rise.

Figure 5 shows typical *I-V* characteristics measured for the D3 detectors. At first glance, these curves look completely different from those measured for the D1 and D2 detectors. The curves have linear dependencies, with only slight diode-like behavior at low biases. Nevertheless, as we describe below, we can in fact use the same physical model for all detector groups. For comparison, Fig. 6 shows two representative curves measured for the D1 and D3 detectors. Because of the small pixel size of the D3 detectors, the measured currents were smaller than those measured for the D1 and D2 detectors at the same bias. This is the reason for the fluctuation seen at low bias for the D3 detectors.

The *I-V* curve measured for D4 detectors are very similar to those measured for D3 and we will discuss them later in conjunction with temperature dependence of dark currents.

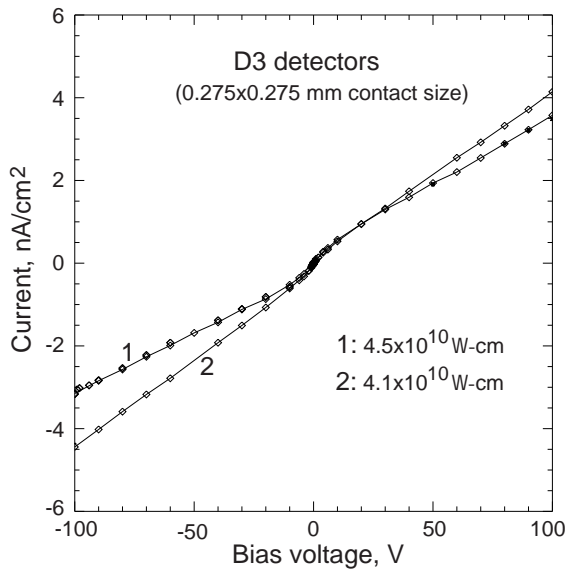

Figure 5. *I-V* characteristic measured for the D3 detectors; the contact size is 400x400 µm; the gap between the contact and the guard ring is 50 µm; the effective contact area used to normalize the current is 0.00012 cm$^2$.

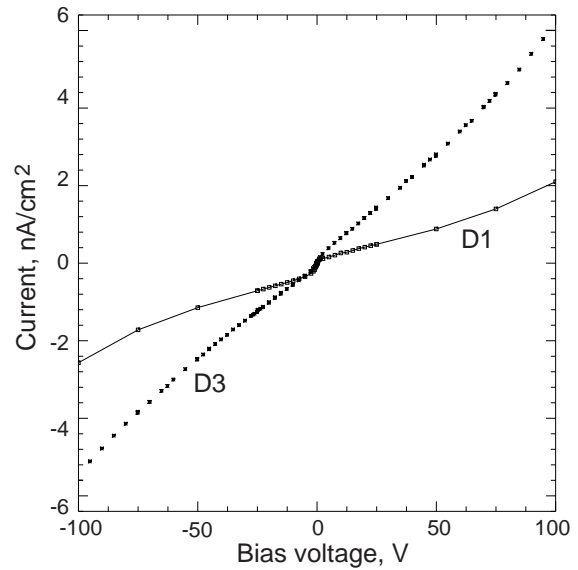

Figure 6. Comparison between representative *I-V* curves measured for the D1 and D3 detectors.

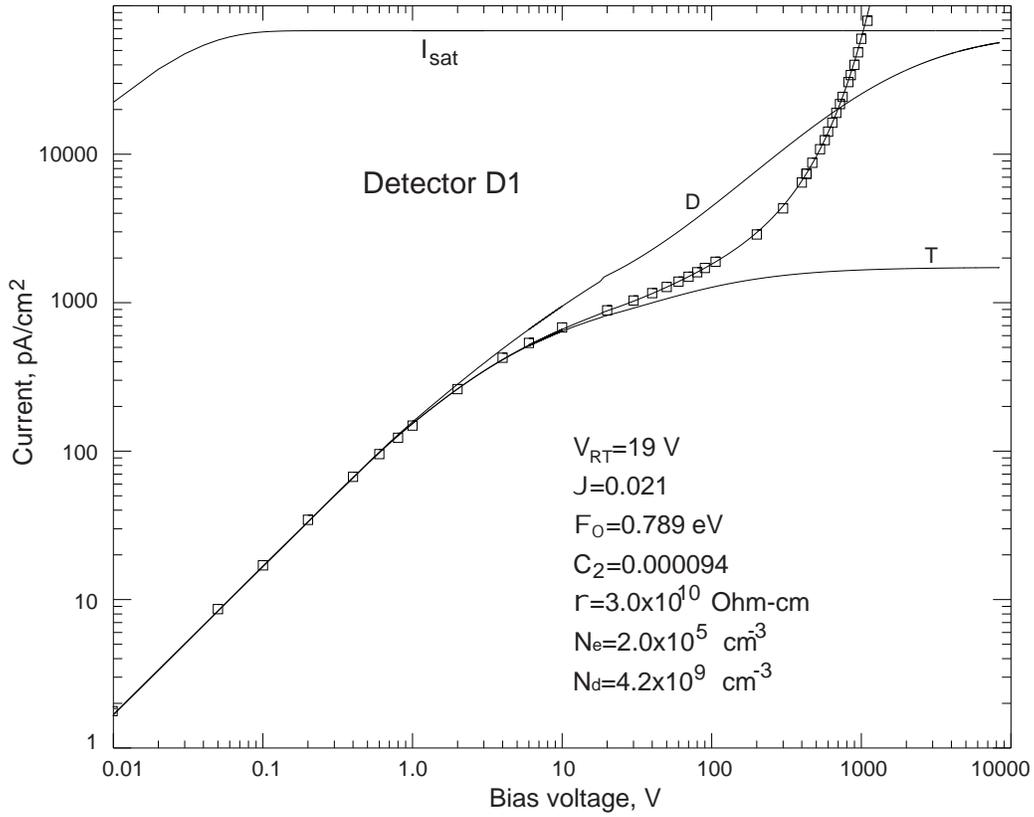

Figure 7. The measured (squares) and calculated (solid lines) *I-V* characteristics of the D1 detector. The curve labeled D is calculated for $\vartheta_n=1$ and $C_2=0$ (no interfacial layer), while the curve labeled T is calculated for $\vartheta_n\neq1$ and $C_2=0$ (no potential barrier lowering). The curve $I_{SAT}$ represents the saturation current of the ideal Schottky barrier in the termionic approximation.

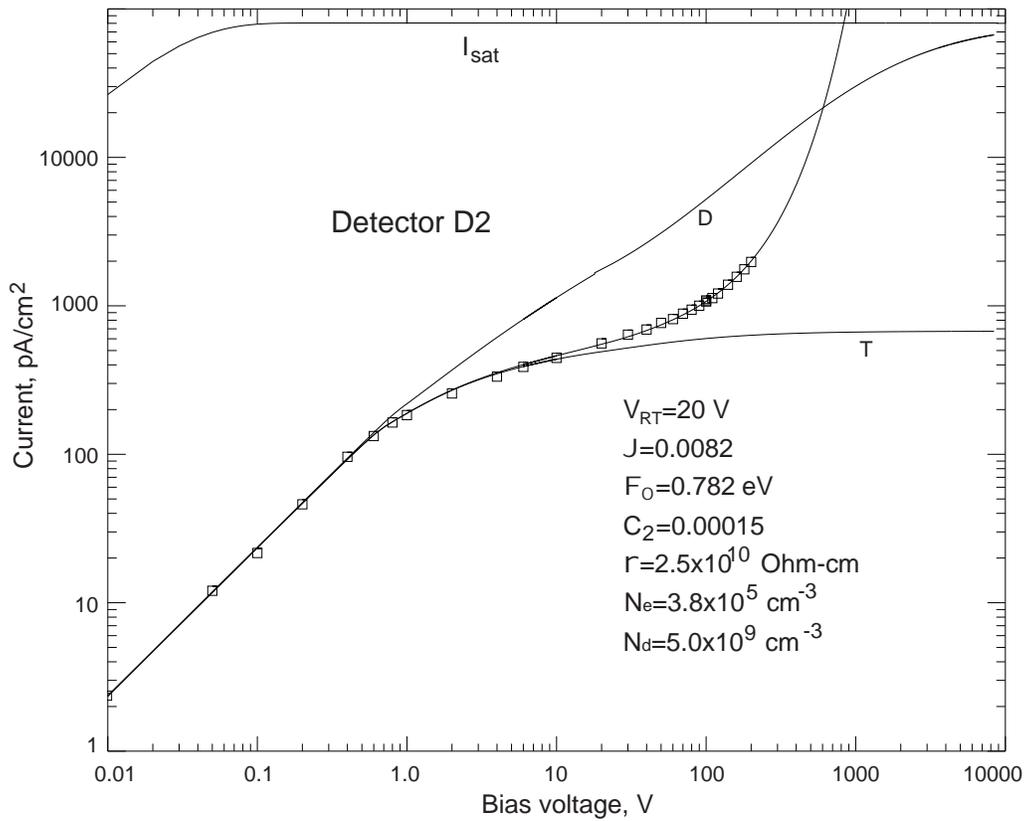

Figure 8. Same as Fig. 7 but plotted for the D2 detector.

We applied the ITD model described in the previous section to fit the measured curves. We found that we can reproduce all the measured *I-V* characteristics accurately. To illustrate the fitting procedures, we selected three representative *I-V* characteristics; a positive branch of the *I-V* curve measured for the D1 detector (large contact); a negative branch measured for the D2 detector (small contacts), and a positive branch measured for the D3 detectors. The experimental curves (squares) and the evaluated theoretical curves (solid lines) are shown in Figs. 7-9 on a log-log scale. Table 1 summarizes the magnitude of the parameters obtained from the least square fit, used to calculate the theoretical curves. As seen, the agreement between the ITD theory and the experimental data is very good. For the D1 curve the $\chi^2$ function has a very broad minimum, and practically any value of $V_{RT}$ between 18 and 70 V provides a satisfactory fit to the data. For the D2 curve the acceptable values of $V_{RT}$ range between 12 and 25 V, with $\chi^2$ reaching the minimum at 19.9 V. Finally, the *I-V* curve measured for the D3 detector gives 9.7 V for $V_{RT}$.

For all groups of the detectors, the corresponding values of $N_D$ were between 0.2 and 2.5 x$10^{10}$ cm$^{-3}$. As seen, the effective concentration of the ionized donors in the depleted volume is much higher than the concentration of the free carriers (electrons) inside the undepleted bulk. This is typical for the highly compensated material. The correlation between the parameters $\rho$, $N$, and $N_D$ is also evident, This is probably related to the total impurity concentration. We found nearly the same barrier heights at zero field for all tested contacts, $\Phi_0$=0.78-0.79 eV, but very different magnitudes of $\vartheta_n$ and $C_2$. Taking 0.8282 eV for the position of the Fermi level inside the CZT bandgap [19], one can find $V_{bi}$~0.03 eV. As seen from Table 1, there is correlation between the parameters $\vartheta_n$ and $C_2$. This can be attributed to the fact that the larger the thickness of the interfacial layer, the smaller the transmission coefficient $\vartheta_n$, and the higher the voltage drop across the interfacial layer ($\Delta V_I = C_2 V$).

The ITD theory allows us to understand the factors determining the bulk leakage currents in the high resistivity CZT detectors. At low voltages, current is always limited by the specific bulk resistivity of CZT, typically 1-5x$10^{10}$ Ohm-cm. In the case of the ideal Schottky barrier, the maximum possible current, $I_{MAX}$, would be equal to the saturation current $I_{SAT}$ across

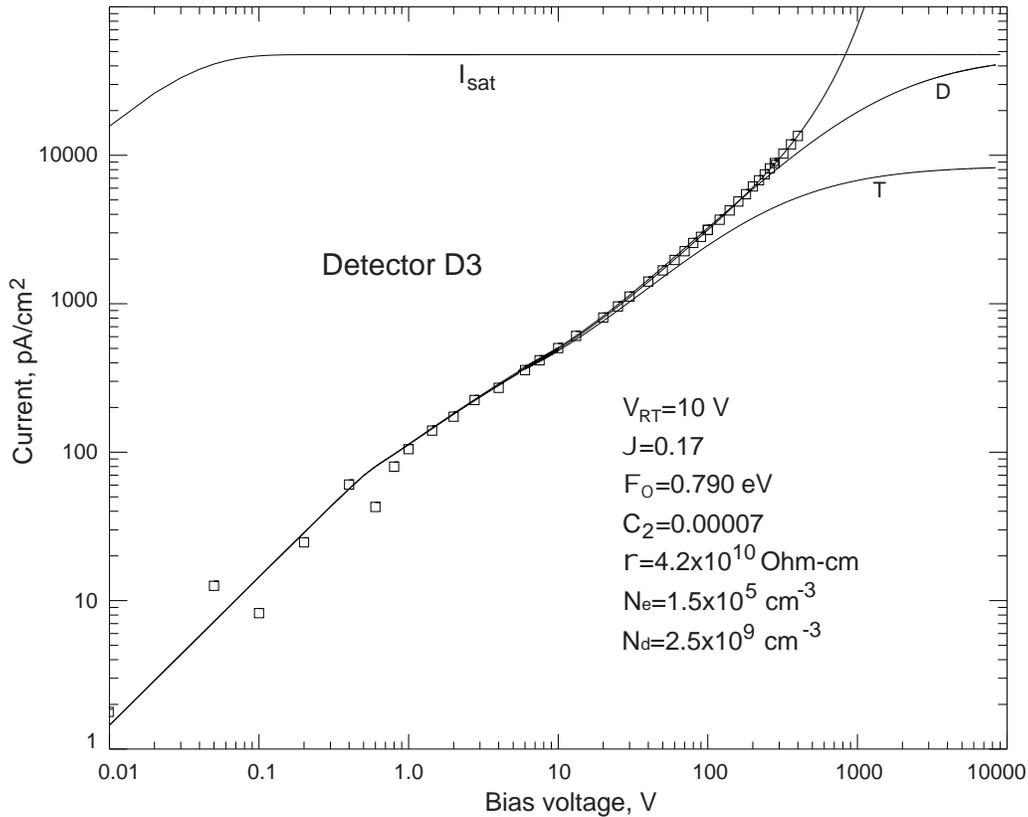

Figure 9. Same as Fig. 7 but plotted for the D3 detector.

the barrier,

$$I_{SAT}=A^*T^2exp(-\Phi_0/V_{TH}). \qquad (21)$$

For comparison, Figs. 7-9 show an ideal Schottky barrier characteristic with the saturation current $I_{SAT}$. If the interfacial layer exists between the contact and semiconductor, the current will be significantly reduced due to the factor $\vartheta_n$ at low biases, and will rise exponentially at very high bias ($\vartheta_n V_R/V_D<<1$) because of the barrier height lowering:

$$I=\vartheta_n I_{SAT}exp(C_2V/V_{TH}). \qquad (22)$$

As an example, for the D1 and D2 detectors, the measured current already exceeds $I_{SAT}$ at biases above >500V. The current $I$, given by Eq. (22), is obtained in the thermionic limit ($\vartheta_n V_R/V_D<<1$), i.e. when all electrons entering the semiconductor are rapidly swept by the electric field. However, if the electron drift velocity $V_D$ is not fast enough to efficiently remove electrons from the near contact area, the resulting current will be smaller. In the diffusion limited current case, i.e. when $\vartheta_n V_R/V_D>>1$, and $V>V_{RT}$, then:

$$I=eN_C\mu E_C exp(-\Phi_0/V_{TH}), \qquad (23)$$

where $E_C$ is the electric field strength at the contact, and $N_C$ is given by Eq. (11). As for the actual current it is hard to say *a priori* if it is thermionic or diffusion-limited. In the general case the current is determined by Eq. (8) from which the diffusion and thermionic limits can be derived, depending on the ratio $\vartheta_n V_R/V_D$.

Table 1

|  | D1, Fig. 7 | D2, Fig. 8 | D3, Fig. 9 | D4, Fig. 10 |
|---|---|---|---|---|
| $\rho$, x$10^{10}$ Ohm–cm | 2.9 | 2.2 | 4.2 | 4.5 |
| $N$, x$10^5$ cm$^{-3}$ | 2.1 | 3.0 | 1.5 | 1.3 |
| $N_D$, x$10^{10}$ cm$^{-3}$ | 0.4-2.7 | 0.5 | 0.25 | 0.25 |
| $V_{RT}$, V | 18-70 | 20 | 9.7 | 12 |
| $\Phi_0$, eV | 0.78-0.79 | 0.782 | 0.790 | 0.788 |
| $\vartheta_n$ | 0.02-0.04 | 0.0082 | 0.17 | 0.12 |
| $C_2$, x$10^{-5}$ | 9.2-9.4 | 15.0 | 6.8 | 6.0 |

To illustrate the effect of the interfacial layer on the dark current, we calculated the theoretical *I-V* curves for two cases: 1) $\vartheta_n=1$ and $C_2=0$, i.e. no interfacial layer, and 2) $\vartheta_n<1$ and $C_2=0$, i.e. no potential barrier lowering. The magnitudes of the remaining parameters were taken from the least square fit of the experimental data. If no interfacial layer exists (first case) the calculated current (curves D in Figs. 7-9) would be diffusion-limited up to very high biases, such that the condition $V_R/V_D>>1$ is satisfied. In other words, the dark current in high resistivity CZT detectors is diffusion-limited if no interfacial layer exists. Eq. (23) can be rewritten as:

$$I=eN_S\mu E_C, \qquad (24)$$

where $N_S$ is the free electron concentration near the contact. On the other hand, in the diffusion approximation the surface concentration $N_S$ can be expressed as:

$$N_S=N_B exp(-V_{bi}/V_{TH}), \qquad (25)$$

where $N_B$ is the free electron concentration in the undepleted bulk. Eq. (24) resembles the Ohmic-like dependence but with a much smaller specific resistivity due to a reduction factor $exp(-V_{bi}/V_{TH})$, e.g. for $V_{bi}=0.05$ V $exp(-V_{bi}/V_{TH})=0.15$. Thus, in the applied bias range from 1 to 100V, the measured *I-V* curve could be misinterpreted as following Ohm's law, and, as was first pointed out in Ref. [4], a significant overestimate of the bulk resistivity would be obtained.

If no potential barrier lowering is assumed, i.e. $C_2=0$, the calculated *I-V* curves, labeled T in Figs. 7-9, would correspond precisely to the termionic-limited current for the detectors D1 and D2, and still be diffusion-limited for D3. As discussed previously, this is why, the *I-V* curves for the detectors from groups D1 and D2 are very different from those measured for D3. It appears that the D1 and D2 detectors have an interfacial layer which makes the condition $\vartheta_n V_R/V_D \ll 1$ exist even at low bias. In contrast, we assume that the D3 detectors have a much thinner layer, with $\vartheta_n \sim 1$, and, as a result the current is diffusion-limited up to high bias.

It is interesting to compare the *I-V* curves measured for D2 (thick interfacial layer) and D3 (thin interfacial layer). Below 1 V the current measured for D3 is approximately 2 times smaller than D2 because of the difference in bulk resistivities: $2.2 \times 10^{10}$ and $4.2 \times 10^{10}$ Ohm-cm. On the contrary, around 200 V, the current measured for D2 becomes 3-4 times smaller, because of the transmission factor $\vartheta_n$, than that measured for D3. At even higher biases, the exponential rise, due to the barrier lowering, dominates, and at some point the D2 current exceeds the D3 current again, as seen in Figs. 7-9. It is clear that for any operating voltage there should be an optimal thickness of the interfacial layer which provides the minimal leakage current. However, the most efficient way to reduce the leakage current is, of course, to use contacts with large barrier heights.

Figure 10 shows the *I-V* characteristics measured for a randomly selected D1 detector at different detector temperatures. We found that the least squares fit for each curve yields similar results within the fitting errors for all parameters of the Schottky barrier. The solid lines represents the theoretical curves calculated after substituting averaged values for the fitting parameters. The temperature dependence of the dark current in the range between 20 to 70 C is shown in Fig. 11 for two cathode biases: 20 and 100 V. The solid line depicts the theoretical curves calculated by using the parameters found from the previous fit shown in Fig. 10.

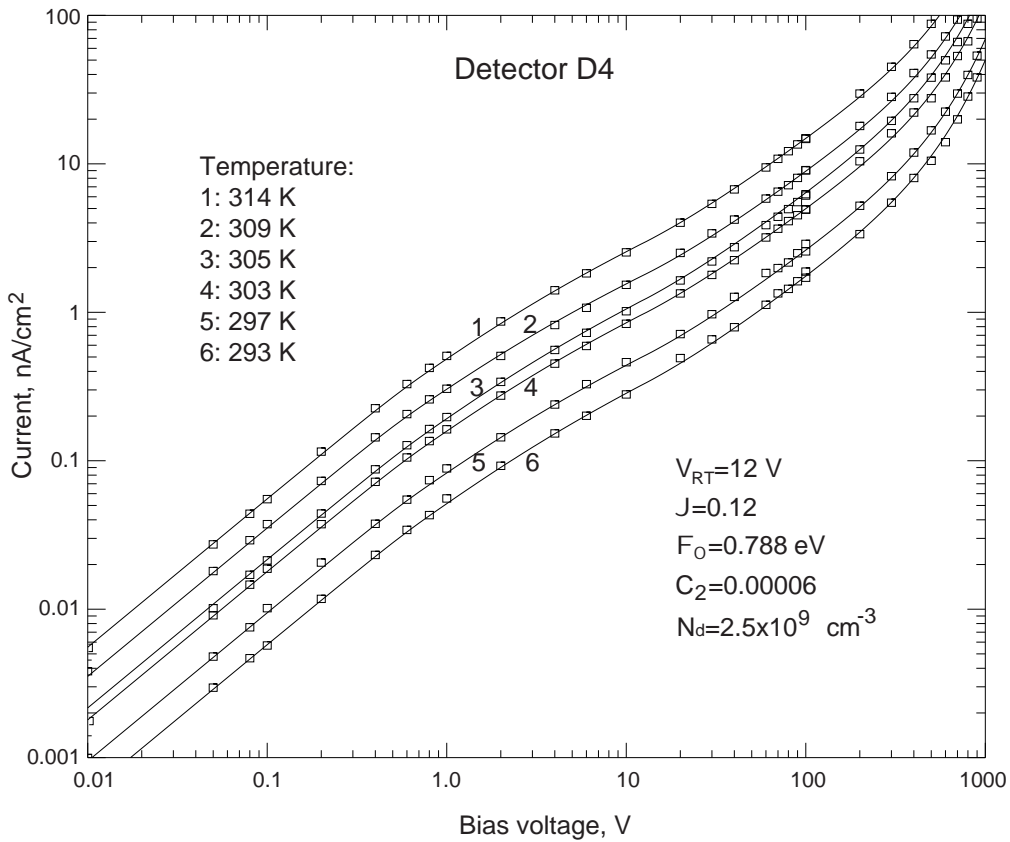

Figure 10. The measured (squares) and calculated (solid lines) *I-V* characteristics of a D4 detector at six detector temperatures. The same set of free parameters was used to calculate the theoretical curves for each temperature.

## 5. Conclusions

We have demonstrated that the bulk *I-V* characteristics measured for the CZT pixel detectors with Pt contacts can be explained by applying a combined interfacial layer-thermionic-diffusion theory to a back-to-back Schottky barrier system. By fitting the measured curves over a 5 decade range we obtain consistent parameters for the Schottky barrier as well as for the CZT material. For example, we found the potential barrier of the Pt contact to be 0.78-0.79 eV.

It appears that the interfacial layer, likely formed during the detector fabrication process, can significantly affect the *I-V* characteristics of CZT detectors with blocking contacts (Pt contacts in this case). The detector leakage current is limited by the material bulk resistivity at low bias (<1V). At high applied voltages, the current is determined by the potential barrier height, transmission coefficient through the interfacial layer, and by the barrier height lowering effect due to the voltage drop across the interfacial layer. If the effect of the interfacial layer is small, the leakage current is diffusion-limited up to very high bias, and can resemble ohmic behavior, with effective bulk resistivity much higher than $5 \times 10^{10}$ Ohm-cm.

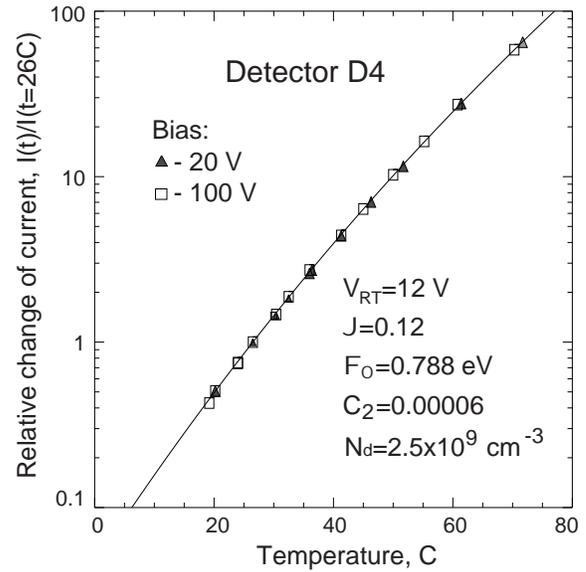

Figure 11. The temperature dependence of the dark current measured (squares) and calculated (solid lines) for 20 and 100 V biases on the cathode.

## Acknowledgments

This work was supported by NASA under grant No. NAG5-5289. The authors wish to thank K. Parhnam and C. Szeles from eV-Products, Inc. for fruitful discussions.